\documentclass[two column, prl]{revtex4}
\usepackage{graphics,graphicx}
\usepackage{ulem}
\begin{document}
\title{Scaling of nonlinear susceptibilities in artificial permalloy honeycomb lattice}
\author{A. ~Dahal}
\author{Y. Chen}
\author{B. Summers}
\author{D. K.~Singh$^{*}$}
\affiliation{Department of Physics and Astronomy, University of Missouri, Columbia, MO 65211}
\affiliation{$^{*}$email: singhdk@missouri.edu}

\begin{abstract}
Two-dimensional artificial magnetic honeycomb lattice is predicted to manifest thermodynamic phase transition to the spin solid order ground state at low temperature. Nonlinear susceptibilities are very sensitive to thermodynamic phase transition. We have performed the analysis of nonlinear susceptibility to explore the thermodynamic nature of spin solid phase transition in artificial honeycomb lattice of ultra-small connected permalloy (Ni$_{0.81}$Fe$_{0.19}$) elements, typical length of $\simeq$ 12 nm. The nonlinear susceptibility, $\chi_{n1}$, is found to exhibit an unusual cross-over character in both temperature and magnetic field. The higher order susceptibility $\chi_3$ changes from positive to negative as the system traverses through the spin solid phase transition at $T_s$ = 29 K. Additionally, the static critical exponents, used to test the scaling of $\chi_{n1}$, do not follow the conventional scaling relation. We conclude that the transition to the ground state is not truly thermodynamic, thus raises doubt about the validity of predicted zero entropy state in the spin solid phase.
\end{abstract}


\maketitle

The interplay between magnetic and thermodynamic characteristics often dictates the nature of phase transition in a magnetic material. Magnetic materials that exhibit equilibrium phase transition, such as spin ice or spin glass, aptly manifest this tendency.\cite{Snyder,Ramirez,Young} More recently, artificial magnetic honeycomb lattice has emerged as new venue to explore many equilibrium phenomena of geometrically frustrated magnets in a disorder-free environment.\cite{Nisoli,Wilis,Tanaka,Ramirez2} The underlying physics in a two-dimensional honeycomb lattice is controlled by the peculiar moment arrangements of 'two-in \& one-out' (or vice-versa) or 'all-in or all-out' configurations on a given vertex of the lattice.\cite{Nisoli,Tanaka} The two-in \& one-out refers to a situation where two moments, aligned along the elements of the honeycomb lattice, are pointing towards the vertex and one moment is pointing away from it; also termed as the quasi-ice rule.\cite{Qi} Theoretical researches have shown that an artificial magnetic honeycomb lattice can undergo a series of thermodynamic phase transitions as a function of reducing temperature from a paramagnetic phase, consisting of the distribution of 'two-in \& one-out' (or vice-versa) and 'all-in or all-out' moment arrangements, to a short-range ordered spin ice state.\cite{Moller,Chern} For further reduction in temperature, the system tends to develop a magnetic charge ordered state, which is described by the random distribution of chiral vortex loops. At much lower temperature, a honeycomb lattice is predicted to develop a novel ground state of spin solid order, described by the periodic arrangements of the vortex magnetic loops of opposite chiralities.\cite{Branford} Each magnetic phase transition reduces the overall entropy of the system. The transition to the spin solid ground state is expected to be truly thermodynamic in nature, with zero entropy and magnetization at low temperature.\cite{Mellado,Zoe,Moller}

Analysis of nonlinear susceptibilities provide an ideal method to test the thermodynamic nature of a magnetic phase transition.\cite{Jonsson,Johnston,Martinez} An equilibrium or thermodynamic phase transition is manifested by the scaling of nonlinear susceptibilities where the static critical exponents are related to each other via a conventional relation. To understand the development of the spin solid order, from thermodynamical point of view, it is desirable to investigate the properties of nonlinear susceptibilities in artificial honeycomb lattice. Previous efforts in accessing the ground state of spin solid order have mostly focused on the disconnected geometry of the honeycomb lattice where thin elements, of length varying between $\simeq$ 500 nm - 2 $\mu$m, are separated enough to reduce the inter-elemental energy of the lattice.\cite{Heyderman1,Heyderman2} More recently, we proposed a new sample design to create artificial honeycomb lattice of 'connected' ultra-small permalloy (Ni$_{0.81}$Fe$_{0.19}$) elements, with typical element dimension of $\simeq$ 12 nm (length) $\times$ 5 nm (width) $\times$ 7 nm (thickness).\cite{Brock} Details about the fabrication procedure can be found elsewhere.\cite{Brock2} At this length scale, the estimated inter-elemental energy, $\simeq$ 12 K, is small enough to allow temperature to be a feasible tuning parameter to explore the temperature dependent evolution of magnetic phases, including the spin solid order. Using magnetic, neutron reflectometry and small angle neutron scattering measurements, previously we demonstrated the phase transition to the long range ordered spin solid state at low temperature, $T \leq$ 30 K, in the newly designed honeycomb lattice.\cite{Brock,Artur} In this letter, we show that the development of spin solid state is accompanied by a change in the nature of nonlinear correction to the linear susceptibility $\chi_1$. As the system traverses through the spin solid transition at $T_s$ $\simeq$ 29 K, the nonlinear term, $\chi_3$, changes from negative to positive, which is atypical of magnetic phase transition. Also, a cross-over between low field and high field regimes is detected, which leads to two different scaling analysis of non-linear susceptibilities. The estimated static critical exponents do not follow the conventional scaling relation. Together, these phenomena suggest that the transition to the spin solid state is not truly thermodynamic.

\begin{figure}
\centering
\includegraphics[width=8.5 cm]{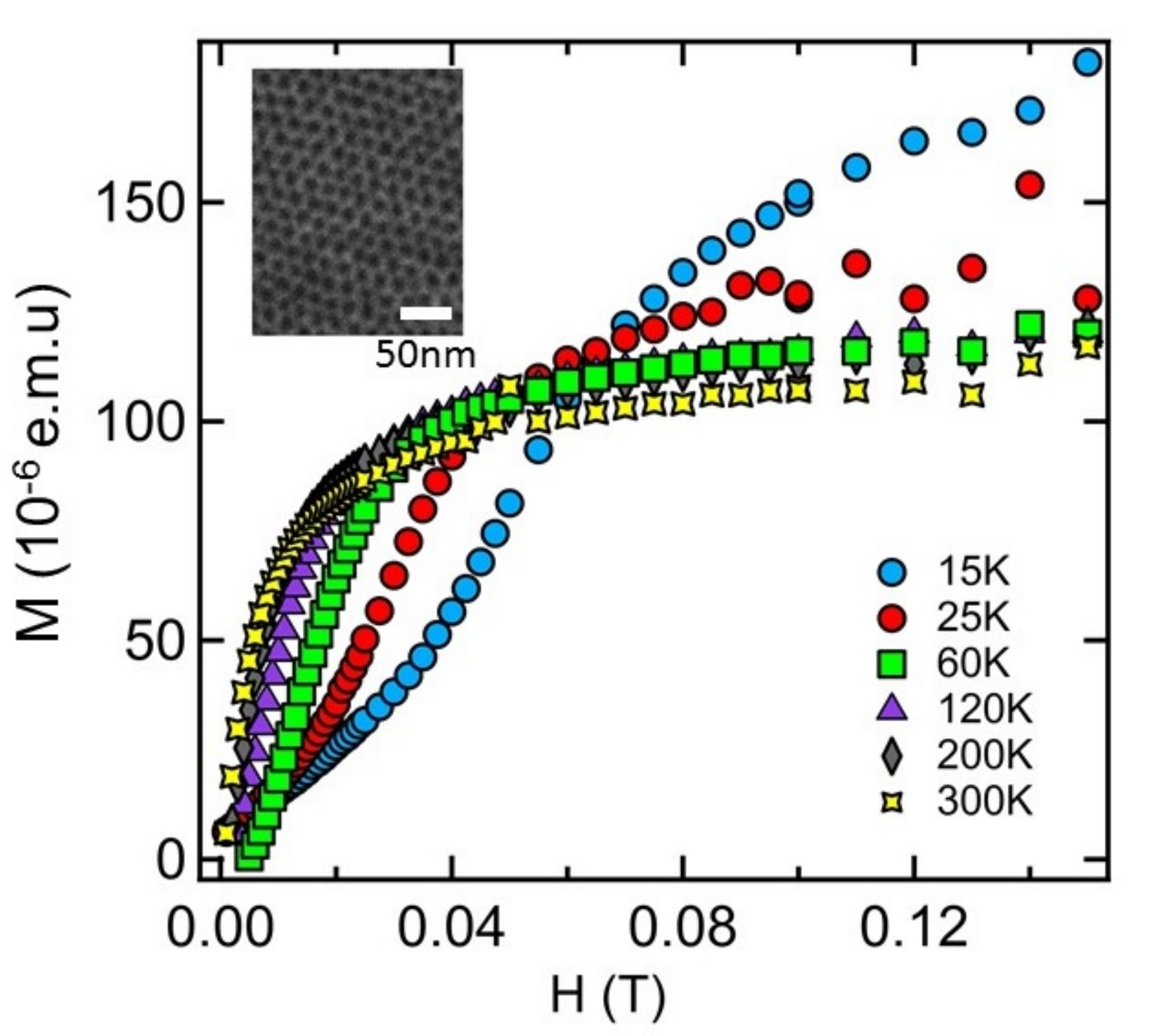} \vspace{-4mm}
\caption{(color online) Magnetization as a function of field. Here, magnetization is plotted as a function of field at different temperatures. Magnetization data exhibits a cross-over behavior in field. While the higher temperature susceptibility is stronger at low field, the magnetization at low temperature is larger above the cross-over field $H \simeq$ 0.04 - 0.06 T. Inset shows the scanning electron micrograph of a typical artificial honeycomb lattice of ultra-small elements. 
} \vspace{-5mm}
\end{figure}

In the case of a thermodynamic transition of equilibrium phenomenon, the nonlinear susceptibilities exhibit a scaling behavior according to the single parameter, given by:\cite{Young,Martinez,Gingras}

\begin{eqnarray}
{\chi}_{n1} ({T}, {H})&=&{H}^{2/{\delta}} {f}({{\tau}^{({\gamma}+{\beta})/2}/H})
\end{eqnarray}

\noindent where $\tau$ = ($T$/$T_s$)- 1, $\gamma$ is the static critical exponent describing the divergent nature of magnetic susceptibility as a function of temperature and $\beta$ is the magnetic order parameter critical exponent. The determination of non-linear susceptibility, $\chi$$_{n1}$, plays the key role in this exercise. The nonlinear susceptibilities are written as the higher order terms in following equations:\cite{Martinez,Gingras}

\begin{eqnarray}
{M}/{H}({T})&=&{\chi_1}({T})-{\chi_3}({T}){H}^{2}+{O}({H}^{4})\\
 &=&{\chi_1}({T})-{a}_{3}({T}){\chi_1}^{3}{H}^{2}+{O}({H}^{4})
\end{eqnarray}
\begin{eqnarray}
{\chi_{n1}}({T}, {H})&=& 1 - {M}({T}, {H})/{\chi_1}{H}
\end{eqnarray}

\noindent where $\chi$$_{1}$(T) is the linear susceptibility at temperature $T$, $\chi$$_{3}$(T) is the nonlinear susceptibility, coefficient $a$$_{3}$ = $\chi$$_{3}$/($\chi$$_{1}$)$^{3}$ and $\chi_{n1}$ is the net nonlinear susceptibility.

\begin{figure}
\centering
\includegraphics[width=8.9 cm]{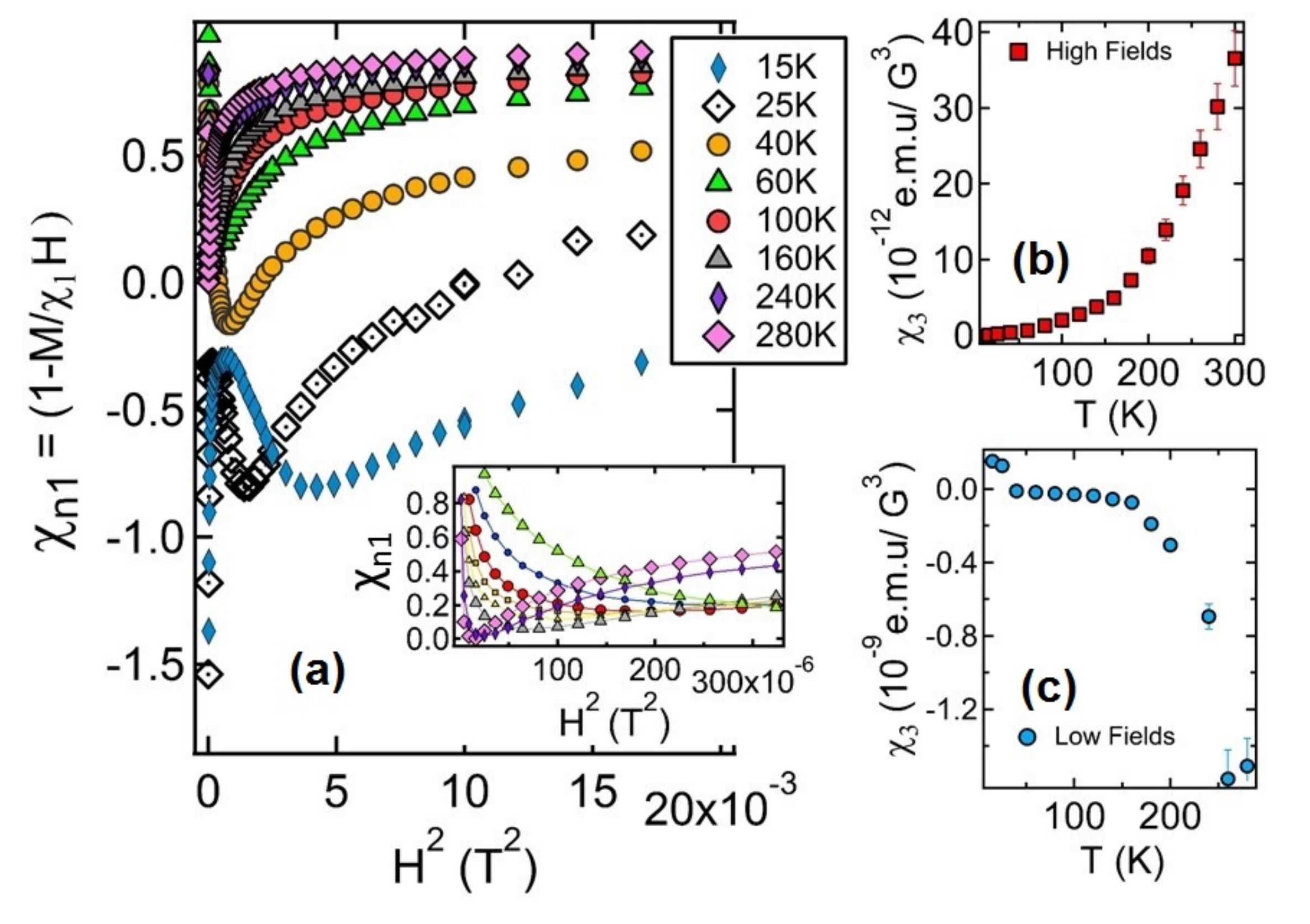} \vspace{-6mm}
\caption{(Color online) Nonlinear susceptibility, $\chi_{n1}$, as functions of field and temperature. (a) $\chi_{n1}$ is estimated using Eq. 2-4 where $\chi_1$  is obtained from fitting $M$ vs $H$ plot at low field. Two features are immediately obvious in this figure: a change in the sign of overall nonlinear susceptibility across $T \simeq$ 30 K and a cross-over regime in field and temperature. As shown in the inset of the figure, the slope of the curve changes from negative to positive at some field value. We call it characteristic cross-over field, which increases as temperature decreases. (b-c) Higher order susceptibility $\chi_3$ as a function of temperature across the cross-over field. $\chi_3$ increases as a function of temperature in high field regime 2 (fig. b) and becomes more negative in low field regime 1 (fig. c). 
} \vspace{-5mm}
\end{figure}

Determination of the critical exponents, $\gamma$ and $\beta$, depends on the asymptotic nature of the arbitrary scaling function $f(x)$, with the boundary conditions $f(x)$ = Constant as $x$$\rightarrow$~0 and $f(x)$ = $x$$^{-2\gamma/(\gamma+\beta)}$ as $x$$\rightarrow$~$\infty$. The nonlinear susceptibility, $\chi_{n1}$($T$, $H$), is expected to follow power-law dependence in both $T$ and $H$ with two independent static critical exponents $\gamma$ and $\delta$, respectively. The power law dependencies are described by the following expressions:\cite{Young,Martinez},

\begin{eqnarray}
{\chi}_{n1}(T)&{\propto}&{\tau}^{|\gamma|}\\
{\chi}_{n1}(T \simeq {T_s, H})&{\propto}&{H}^{2/\delta}
\end{eqnarray}

\noindent The two independent exponents, $\gamma$ and $\delta$, are related to the magnetic order parameter critical exponent $\beta$ via the following scaling relation:

\begin{eqnarray}
{|\delta|}&=& 1+{|\gamma}/{\beta|}
\end{eqnarray}

\noindent The above scaling relation represents a robust test, arguably, of the true equilibrium (thermodynamic) phase transition in a magnetic system. Magnetization data on the newly designed artificial permalloy (Ni$_{0.81}$Fe$_{0.19}$) honeycomb lattice were obtained in the field range of 10 - 1500 Oe using a commercial magnetometer. The sample was slowly cooled from $T$ = 350 K to the desired temperature before collecting the data. Extra care was taken in removing magnetic hysteresis in the superconducting magnet of the magnetometer by cycling the magnetic field in oscillatory mode several times at $T$ = 350 K before cooling to the measurement temperature. In Fig. 1, we plot the $M$ vs $H$ data at few characteristic temperatures. The total magnetization at higher temperature is stronger at low field. The trend reverses across the cross-over field, which also varies with temperature. The linear susceptibility, $\chi_{1}$($T$), at different temperatures were determined by fitting the $M$ versus $H$ curves at low fields, see Fig. S1 in the Supplementary Materials. We have analyzed first and second order term in the magnetization data. Beyond the second order term, the non-linear susceptibility becomes much smaller to be of any quantitative importance. Therefore, equation (2) reduces to $\chi_3$($T$, $H$) $H$$^{2}$= 1 - $M(T, H)$/$\chi_1$$H$. Hence, $\chi_{n1}$($T$, $H$) becomes ($\chi_3$/$\chi_1$) ($T$, $H$)$H$$^{2}$.\cite{Monod}

In Fig. 2a, we have plotted net nonlinear susceptibilities, $\chi_{n1}$($T$, $H$), as a function of $H^2$ at different temperatures between $T$ = 10 K and $T$ = 300 K. The plot of nonlinear susceptibility reveals several very interesting behaviors in applied field. First, at low temperature, $T \leq$ 25 K, $\chi$$_{n1}$ is negative for the entire field application range. The negative nonlinear susceptibility suggests that the higher order correction to the linear susceptibility is very strong. Surprisingly, negative $\chi_{n1}$ is only observed below the spin solid phase transition. Second, the nonlinear susceptibility not only becomes positive above $T \simeq$ 30 K, but also exhibits an unusual trend at low field. At low field, $\chi$$_{n1}$ first decreases before manifesting a gradual enhancement as the applied field strength increases. Thus, the slope of the curve changes from negative (regime 1) in low field to positive (regime 2) in high field. Additionally, the slope of the curve also changes as a function of temperature at low field: from positive at $T \leq$ 30 K to negative at $T \geq$ 30 K. We summarize these observations in plot of $\chi_3$ vs. $T$ in different field regimes in Fig. 2b-c. In general, nonlinear correction to the susceptibility only changes in magnitude, not in sign. This is a puzzling behavior in artificial honeycomb lattice. The characteristic cross-over field, separating the two distinct regimes, decreases as the measurement temperature increases (see inset in Fig. 2a). We also notice that the saturated value of $\chi_{n1}$ increases as temperature increases. The net magnetization is expected to decrease as temperature reduces in artificial honeycomb lattice. First, we analyze the non-linear susceptibility data above the characteristic field (in regime 2). Even in regime 2, the maximum value of the field, up to which $\chi$$_{n1}$(T) is linear in $H^2$, decreases gradually as $T$ approaches $T_s$. It suggests that the higher order corrections in the net susceptibility is still significant.\cite{Monod} The linear portion of $\chi$$_{n1}$(T) at different temperatures are fitted with Eq. (3) to extract the coefficient $a$$_{3}$(T). 

\begin{figure}
\centering
\includegraphics[width=8.9 cm]{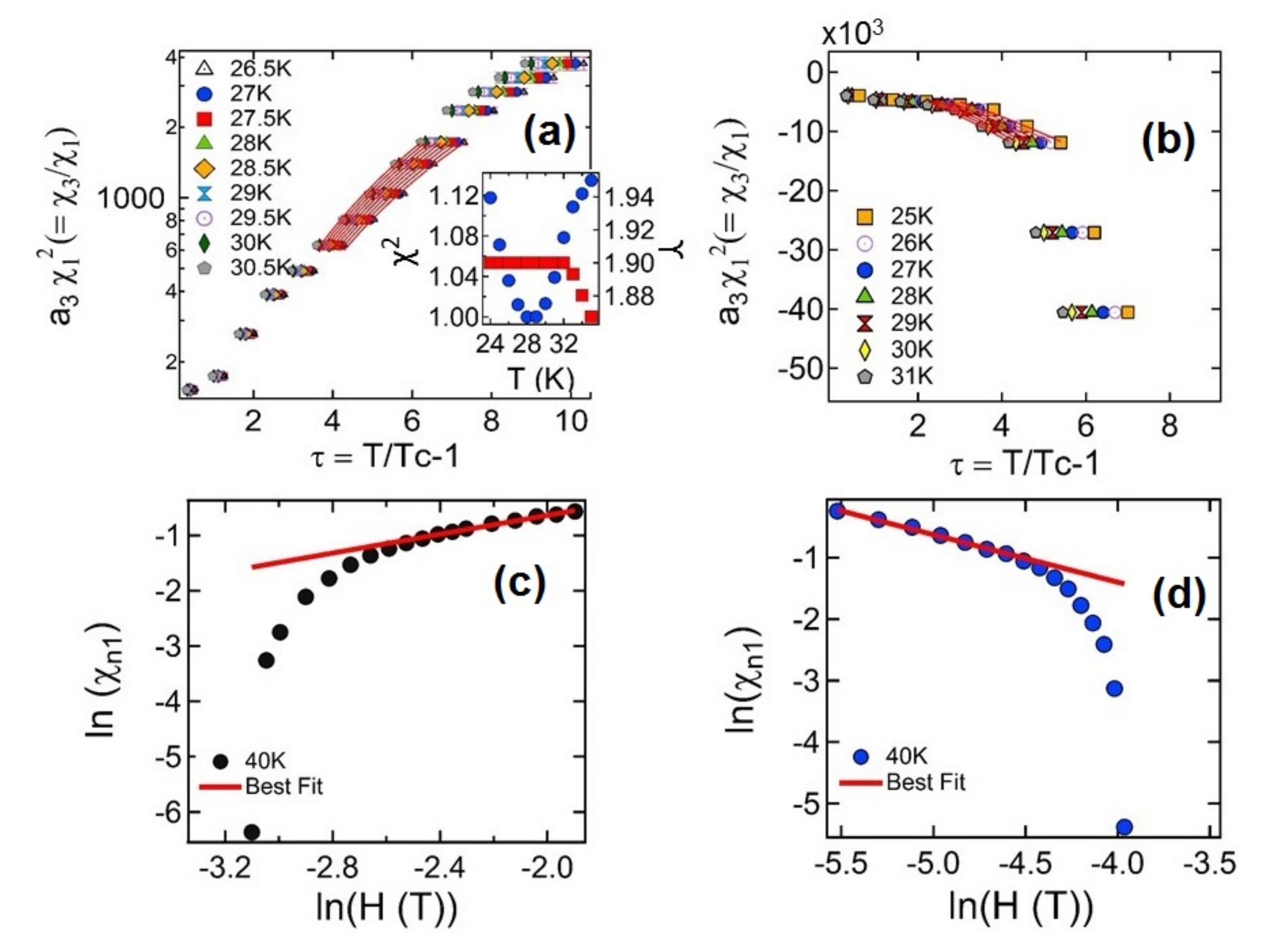} \vspace{-6mm}
\caption{(color online) Estimation of static critical exponents $\gamma$ and $\delta$. (a) To estimate the critical exponent $\gamma$, the coefficient $a_3$ (see text for detail) is plotted as a function of $\tau$ = ($T$/$T_s$)- 1 for different $T_s$ values, across the spin solid transition at $T$ = 30 K. $\gamma$ is estimated by fitting the fixed number of points in the divergence regime of the curve using eq. 5. Best fit to the experimental data is obtained for the critical exponent $|\gamma|$ = 1.9 (inset shows the plot of fitting parameter $\chi^2$ vs $\gamma$). (b) Similar analysis is performed for the low field regime 1, with estimated $|\gamma|$ = 1.4. In both regimes, best fit corresponds to spin solid transition at $T_s$ = 29 K. Nonlinear susceptibility $\chi_{n1}$ is plotted as a function of field at temperature near $T_s$. Experimental data is fitted using the asymptotic function in eq. 6 to obtain critical exponent $|\delta|$ (c) in high field regime 2, $\simeq$ 2.4 and (d) low field regime 1, $\simeq$ 2.5.
} \vspace{-6mm}
\end{figure}

To verify the thermodynamic nature of magnetic phase transition to the spin solid state, first we extract the exponent $\gamma$ using the formalism, described above, in eq. (5). For this purpose, the nonlinear susceptibility $\chi_{n1}$ = $a_3$$\chi$$_{1}$$^{2}$ is plotted as a function of $\tau$ for few different choices of spin solid transition temperatures $T_s$ $\in$ [25, 35] K in Fig. 3a. We have fitted a fixed number of data points, in the divergence regime, on each curve using eq.(5). Estimated $\gamma$ is found to vary in the range of [1.7, 2]. The best fit is obtained for $T_s$ = 29 K, with the corresponding value of $|\gamma|$ = 1.9 (see inset in Fig. 3a). The transition temperature, $T$$_{s}$, is very close to the experimental value of $T$ = 30 K, as estimated from the previous dc susceptibility and electrical measurements.\cite{Brock,Brock2} Also, the static critical exponent $\gamma$ is comparable to the value ($|\gamma|$ $\simeq$ 2.25) found in systems manifesting truly thermodynamic phase transition, such as interacting arrays of nano islands or spin freezing in canonical and geometrically frustrated systems.\cite{Jonsson,Gingras,Martinez,Bouchiat,Young} Similar analysis was performed in the low field regime (regime 1) below the characteristic cross-over field. The best fit is obtained for the static critical exponent $|\gamma|$ = 1.4, see Fig. 3b. It is not very different from the magnitude of $\gamma$ in the high field regime (regime 2). It seems that the cross-over phenomenon, manifested by the change in the slope of $\chi$$_{n1}$(T) as the system traverses across the transition temperature at a given field, does not affect the estimation of $\gamma$ and the transition temperature, $T_s$, in the honeycomb lattice of ultra-small elements.

Next, we determine another critical exponent $\delta$ by plotting ln($\chi$$_{n1}$) versus ln($H$) at temperature near the spin solid transition. The experimental data is fitted using the asymptotic function in eq. (6). As shown in Fig. 3c, a good fit to the data is obtained for the critical exponent $\delta$ = 2.4 in regime 2. Similar analysis in regime 1 at low field yields $|\delta|$ = 2.5, which is also similar in magnitude as found in the high field regime 2. Finally, we test the scaling behavior of non-linear susceptibilities, as described by equation (1). If the magnetic phase transition to the spin solid state in artificial honeycomb lattice is indeed a true equilibrium phase transition, then the nonlinear susceptibilities should exhibit the scaling behavior due to the estimated critical exponents. According to equation 7, for critical coefficients $|\gamma|$ = 1.9 and $\delta$ = 2.4, the magnetic order parameter critical exponent $\beta$ is $\simeq$ 1.4. As shown in Fig. S2 in the Supplementary Materials, the nonlinear susceptibilities at different temperatures do not exhibit the scaling collapse on one curve for the estimated exponents. To explore the scaling behavior further, we vary the critical exponents, $\gamma$, $\delta$ and $\beta$, systematically. First we discuss the scaling in regime 2. A scaling behavior is observed for exponents $\delta$ = 10 and $|\gamma|$ = 1.5, see Fig. 4a. Although exponent $|\gamma|$ is similar to the estimated value, scaling collapse of $\chi_{n1}$ data only occurs for $\delta$ much larger than the estimated value. At large $x$ values, some data scatter from the scaling curve due to the large errors associated with the smaller nonlinear susceptibilities. We also tested the scaling behavior for intermediate values of $\delta$, 4.75, while keeping the coefficient $\gamma$ constant. The scaling of non-linear susceptibilities improves as $\delta$ increases. However, the critical exponents do no follow the scaling relation, outlined in eq. 7. 

\begin{figure}
\centering
\includegraphics[width=8.8 cm]{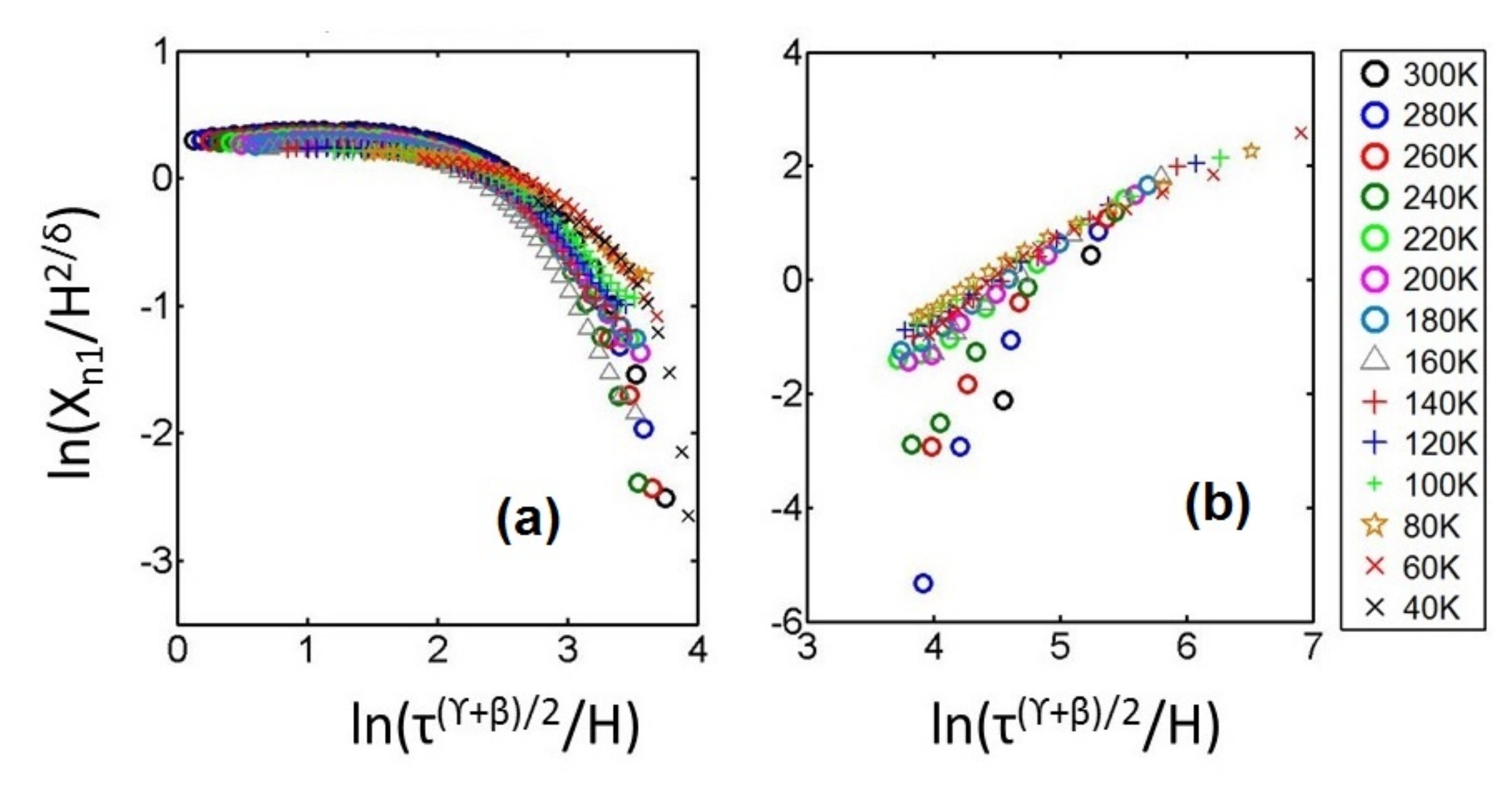} \vspace{-6mm}
\caption{(Color online) Scaling analysis of nonlinear susceptibilities in artificial honeycomb lattice. (a) Nonlinear susceptibilities exhibit scaling behavior for $|\gamma|$ = 1.5, $\delta$ = 10 and $|\beta|$ = 0.1. The critical scaling coefficients do not satisfy the scaling relation in eq. (7). (b) Similar analysis was performed in the low field regime 1. Interestingly, the nonlinear susceptibilities exhibit scaling behavior for the same set of critical exponents, as in high field regime 2. 
} \vspace{-7mm}
\end{figure}

The scaling behavior was also tested for nonlinear susceptibilities in low field regime 1. For uniformity, we have used the estimated static critical exponents of $|\gamma|$ = 1.4, $\delta$ = 2.5 and $|\beta|$ = 0.95 for the scaling analysis. As shown in Fig. S3 of the Supplementary Materials, the non-linear susceptibilities do not scale for the calculated values of exponents. To our surprise, $\chi_{n1}$ data at different temperature exhibit scaling characteristic for the similar set of exponents, $|\gamma|$ = 1.4, $\delta$ = 10 and $|\beta|$ = 0.1, that are used to obtain scaling collapse in the high field regime 2, see Fig. 4b. Once again, the critical exponents do not satisfy the scaling relation in eq. (7). It further confirms that the magnetic phase transition to the spin solid state is not thermodynamic in nature. The observed consistencies in the estimation of critical exponents as well as in the scaling analysis in two different regimes of $\chi_{n1}$ constitute a unique aspect of the spin solid phase transition. It suggests that the nonlinear correction to magnetic susceptibility in spin solid phase is subtly similar to that in the high temperature phases. The discrepancies between the estimated values of the static critical exponents and that used for the scaling manifestation can be attributed, arguably, to the formation of small ferromagnetic clusters with short-range order at intermediate temperatures, which ultimately enhances $\chi$$_{n1}$ considerably and led to strong but non-critical background temperature dependence. Similar behavior was previously observed in magnetic systems that exhibit non-equilibrium phase transition.\cite{Binder}

Our investigation of the thermodynamic nature of magnetic phase transition in artificial honeycomb lattice has revealed two important properties that are not conventional in nature: first, the nonlinear susceptibility exhibits a cross-over behavior in both temperature and magnetic field. The slope of $\chi_{n1}$, which is used to determine the strength of the non-linear correction to the overall magnetic susceptibility, is found to change from negative, at low field, to positive, at high field. Also, the net nonlinear susceptibility, $\chi_{n1}$, changes from positive to negative in temperature. This cross-over occurs across the spin solid phase transition temperature at $T \simeq$ 30 K. A magnetic phase transition is not known to depict such contrasting characteristic across the transition temperature. Clearly, the underlying magnetism in artificial honeycomb lattice does not fit congruently with the conventional understanding. Second, the experimental data do not exhibit scaling behavior for the estimated values of critical exponents. Rather, a scaling collapse of $\chi_{n1}$ requires much larger value of the critical exponent $\delta$; not typically observed in a magnetic material with equilibrium phase transition. Also, the static critical exponents do not satisfy the conventional thermodynamic scaling relation. The overall scaling behavior suggests a non-conventional nature of the transition, which can be arising either due to the finite spin dynamics in the system or, a distribution of relaxation times in short-range ordered magnetic clusters, such as spin ice order or the vortex loop type magnetic correlation across one honeycomb. A distribution of relaxation times in magnetic clusters is known to cause non-conventional scaling behavior. The presence of spin dynamics or the distribution in spin relaxation rate, especially at low temperature, will result in finite entropy accumulation. An artificial magnetic honeycomb lattice is predicted to manifest a zero entropy spin order at low temperature. Previously, researchers devised a statistical method to directly estimate the entropy in artificial two-dimensional frustrated geometry.\cite{Lammert} However, the zero entropy in the spin solid phase is not verified in either geometry, connected or disconnected elements, of the honeycomb lattice. Further research works are highly desirable to fully understand the perplexing observations reported here.

The research at MU was supported by the U.S. Department of Energy, Office of Basic Energy Sciences under Grant No. DE-SC0014461.

\clearpage


\begin{thebibliography}{99}


\bibitem{Snyder} J. Snyder, J. Slusky, R. J. Cava and P. Schiffer, \textit{Nature} \textbf{413}, 48 (2001)

\bibitem{Ramirez} A. P. Ramirez \emph{et al.}, \textit{Nature} \textbf{399}, 333 (1999)

\bibitem{Young} K. Binder and A. P. Young, \textit{Rev. Mod. Phys} \textbf{58}, 801 (1986)

\bibitem{Nisoli} C. Nisoli, R. Moessner and P. Schiffer, \textit{Rev. Mod. Phys.} \textbf{85}, 1473 (2013).

\bibitem{Wilis} A. S. Wills, R. Ballou and C. Lacroix, \textit{Phys. Rev. B} \textbf{66}, 144407 (2002)

\bibitem{Tanaka} M. Tanaka, E. Saitoh, H. Miyajima, T. Yamaoka, and Y. Iye, \textit{Phys. Rev. B} \textbf{73}, 052411 (2006)

\bibitem{Ramirez2} R. Moessner and A. P. Ramirez, \textit{Physics Today}, Feb 2006.

\bibitem{Qi} Y. Qi, T. Brintlinger, and J. Cumings, \textit{Phys. Rev. B} \textbf{77}, 094418 (2008)

\bibitem{Moller} G. Moller and R. Moessner, \textit{Phys. Rev. B} \textbf{80}, 140409 (R) (2009)

\bibitem{Chern} G. W. Chern, P. Mellado, and O. Tchernyshyov, \textit{Phys. Rev. Lett.} \textbf{106}, 207202 (2011)

\bibitem{Branford} W. R. Branford, S. Ladak, D. E. Read, K. Zeissler, and L. F. Cohen, \textit{Science} \textbf{335}, 1597 (2012)

\bibitem{Mellado} P. Mellado, O. Petrova, Y. Shen and O. Tchernyshyov, \textit{Phys. Rev. Lett.} \textbf{105}, 187206 (2010)

\bibitem{Zoe} Z. Budrikis, P. Politi, and R. L. Stamps, \textit{Phys. Rev. Lett.} \textbf{107}, 217204 (2011)

\bibitem{Jonsson} T. Jonsson, P. Svedlindh and M. F. Hansen, \textit{Phys. Rev. Lett.} \textbf{81}, 3976 (1998)

\bibitem{Johnston} D. C. Johnston, \textit{Phys. Rev. Lett.} \textbf{62}, 957 (1989)

\bibitem{Martinez} B. Martinez, A. Labarta, R. Rodríguez-Solá, X. Obradors, \textit{Phys. Rev. B} \textbf{50}, 15779 (1994)

\bibitem{Heyderman1} O. Sendetskyi \emph{et al.}, \textit{Phys. Rev. B} \textbf{93}, 224413 (2016)

\bibitem{Heyderman2} L. Anghinolfi \emph{et al.}, \textit{Nature Comm.} \textbf{6}, 8278 (2015)

\bibitem{Brock} B. Summers, Y. Chen, A. Dahal and D. K. Singh, \textit{Scientific Reports} \textbf{7}: 16080 (2017); DOI:10.1038/s41598-017-15786-8.

\bibitem{Brock2} B. Summers, L. Debeer-Schmitt, A. Dahal, A. Glavic, P. Kampschroeder, J. Gunasekera and D. K. Singh, (under review)

\bibitem{Artur} A. Glavic, B. Summers, A. Dahal, J. Kline, A. Sukhov, A. Ernst and D. K. Singh, \textit{Advanced Science} (Accepted, in press) (2017) 

\bibitem{Gingras} M. J. P. Gingras, C. V. Stager, N. P. Raju, B. D. Gaulin, and J. E. Greedan, \textit{Phys. Rev. Lett.} \textbf{78}, 947 (1997)

\bibitem{Monod} P. Monod \emph{et al.}, \textit{J. Phys. (Paris) Lett} \textbf{43}, 145 (1982)

\bibitem{Bouchiat} H. Bouchiat, \textit{J. Phys. (Paris) Lett} \textbf{47}, 71 (1986)

\bibitem{Binder} K. Binder, \textit{Z. Phys. B} \textbf{48}, 319 (1982)

\bibitem{Lammert} P. E. Lammert, X. Ke, Jie Li, C. Nisoli, D. M. Garand, V. H. Crespi and P. Schiffer, \textit{Nature Physics} \textbf{6}, 786 (2010)








\end{thebibliography}
\end{document}